\documentclass[twocolumn]{aastex62}
\pdfoutput=1 %for arXiv submission
\usepackage{amsmath,amstext,amssymb}
\usepackage[T1]{fontenc}
\usepackage{apjfonts} 
\usepackage{subfigure}
\usepackage{xcolor}
\usepackage{soul}
\usepackage{comment}
\usepackage{graphicx}
\setstcolor{red}

\graphicspath{{figures/}}

\definecolor{britishracinggreen}{rgb}{0.0, 0.42, 0.24}

\newcommand{\cosvar}{\Vec{\theta}}

\shorttitle{Do LIGO/Virgo black hole mergers produce AGN flares?}
\shortauthors{A.~Palmese, M.~Fishbach, et al.}

\begin{document}

\title{Do LIGO/Virgo black hole mergers produce AGN flares? \\
The case of GW190521 and prospects for reaching a confident association}

\author[0000-0002-6011-0530]{A.~Palmese}\affiliation{Fermi National Accelerator Laboratory, P. O. Box 500, Batavia, IL 60510, USA}
\affiliation{Kavli Institute for Cosmological Physics, University of Chicago, Chicago, IL 60637, USA}
\correspondingauthor{Antonella Palmese}
\email{palmese@fnal.gov}
\author[0000-0002-1980-5293]{M.~ Fishbach}\altaffiliation{NASA Hubble Fellowship Program Einstein Postdoctoral Fellow}\affiliation{Center for Interdisciplinary Exploration and Research in Astrophysics (CIERA) and Department of Physics and Astronomy,
Northwestern University, 1800 Sherman Ave, Evanston, IL 60201, USA}
\author[0000-0001-9947-6911]{C.~J.~Burke}\affiliation{Department of Astronomy, University of Illinois at Urbana-Champaign, Urbana, IL 61801, USA}\affiliation{National Center for Supercomputing Applications, University of Illinois at Urbana-Champaign, Urbana, IL 61801, USA}
\author[0000-0002-0609-3987]{J.~T.~Annis}\affiliation{Fermi National Accelerator Laboratory, P. O. Box 500, Batavia, IL 60510, USA}
\author[0000-0003-0049-5210]{X.~Liu}\affiliation{Department of Astronomy, University of Illinois at Urbana-Champaign, Urbana, IL 61801, USA}\affiliation{National Center for Supercomputing Applications, University of Illinois at Urbana-Champaign, Urbana, IL 61801, USA}

%\affiliation{}
%\author{others}

\begin{abstract}

The recent report of an association of the gravitational-wave (GW) binary black hole (BBH) merger GW190521 with a flare in the Active Galactic Nuclei (AGN) J124942.3+344929 has generated tremendous excitement. However, GW190521 has one of the largest localization volumes amongst all of the GW events detected so far. The 90\% localization volume likely contains $7,400$ unobscured AGN brighter than $g \leq 20.5$ AB mag, and it results in a $\gtrsim 70\%$ probability of chance coincidence for an AGN flare consistent with the GW event. We present a Bayesian formalism to estimate the confidence of an AGN association by analyzing a population of BBH events with dedicated follow-up observations.
Depending on the fraction of BBH arising from AGNs, counterpart searches of 
$\mathcal{O}(1)-\mathcal{O}(100)$ GW events  are needed to establish a confident 
association, and more than an order of magnitude more for searches without followup (i.e, using only the locations of AGNs and GW events). Follow-up campaigns of the top $\sim 5\%$ (based on volume localization and binary mass) of BBH events with total rest frame mass $\ge 50~M_\odot$ are expected to establish a confident association during the next LIGO/Virgo/KAGRA observing run (O4), as long as the true value of the fraction of BBH giving rise to AGN flares is $>0.1$. Our formalism allows us to jointly infer cosmological parameters from a sample of BBH events that include chance coincidence flares. 
Until the confidence of AGN associations is established, the probability of chance coincidence must be taken into account to avoid biasing astrophysical and cosmological constraints.

\end{abstract}


\reportnum{FERMILAB-PUB-21-020-AE}

\section{Introduction}

One of the most interesting gravitational wave (GW) detections to date is the binary black hole merger GW190521 \citep{GW190521}. This event is associated with the most massive binary system detected by LIGO/Virgo so far, with a total mass of $\sim 150~M_\odot$. This makes GW190521 particularly interesting, since the origin of black holes in the mass gap challenges the standard theories of stellar evolution \citep{190521_properties}, although the origin of this event as isolated binary cannot be excluded \citep{farrell2020gw190521,kinugawa2020formation}, and the components mass may fall outside of the mass gap \citep{fishbach2020dont}. This detection therefore resulted in a large number of proposed alternative formation scenarios including primordial black holes \citep{pbh_190521}, exotic Proca stars~\citep{PhysRevLett.126.081101}, low--mass dwarf galaxy mergers \citep{Conselice_20,palmese_conselice}, dynamical interactions in dense stellar environments \citep{romeroshaw2020gw190521,gayathri2020gw190521,fragione2020origin} and black holes grown by accretion \citep{safarzadeh2020formation}. The latter scenario can also occur in Active Galactic Nuclei (AGN) disks, although the accretion probably happens at a relatively low rate, affecting BH masses by $\lesssim 10\%$ \citep{Tagawa,2020ApJ...901L..34Y}.
A compelling explanation for the formation of massive stellar black holes is through repeated mergers of smaller black holes~\citep{2017ApJ...840L..24F,2017PhRvD..95l4046G}, and such hierarchical mergers are a natural prediction for BBHs assembled in AGN disks~\citep{2019PhRvL.123r1101Y}. Because of the gas-rich environment, BBH mergers in AGN disks may also give rise to electromagnetic counterparts through several mechanisms (e.g. \citealt{McKernan12,Bartos_2017,McKernan19,kimura2021}). This is particularly relevant for GW190521 because \citet{Graham} (G20) found a potential electromagnetic (EM) counterpart in AGN J124942.3+344929 using Zwicky Transient Facility (ZTF; \citealt{Masci_2018,Bellm_2018}) observations. 

The prospect of EM counterparts to BBH events is exciting for several reasons, including the potential for standard siren cosmology~\citep{schutz,2005ApJ...629...15H,chen17,palmese19}. GW events have already been used to independently measure the Hubble constant (\citealt{2017Natur.551...85A,fishbach,2019arXiv190806060T}; \citealt*{darksiren1}; \citealt{Palmese:2020aof}). Meanwhile, the AGN association to GW190521 has also been used in several works to derive cosmological constraints \citep{chen2020standard,gayathri2020standard,mukherjee2020standard,2020RNAAS...4..209H}. However, these analyses do not account for the probability of a chance coincidence, which is particularly significant because GW190521 has the second largest localization in terms of comoving volume encompassed amongst all GW detection so far (see Table \ref{tab:vol} for the volume of a selected sample of LIGO/Virgo events). Moreover, \citet{depaolis} showed that this AGN flare can also be explained by a microlensing event.

In this work, we expand on the analysis presented in \citet{ashton2020current} and find insufficient evidence for a common origin for GW190521 and the AGN flare. We explore the uniqueness of the candidate and the odds of chance coincidence for similar flares based on the population of AGNs expected in the entire and observed GW190521 localization volumes. We then turn to a population of GW events with possible AGN counterparts, and define a Bayesian formalism that allows us to derive the number of GW events needed to establish a confident association between GW BBH events and AGN flares. This problem was first explored in ~\citet{Bartos} (hereafter B17), but here we consider GW events with targeted followup observations to catch transients, rather than an existing catalog of AGN locations. This statistical framework is presented in Section \ref{sec:method}.
In Section \ref{sec:190521}, we present results for the case of GW190521. Section \ref{sec:future} presents prospects for making confident associations in the future by measuring the fraction of BBH events that induce AGN flares and Section \ref{sec:futurecosmo} presents prospects for simultaneously using GW and AGN observations for standard siren cosmology. We conclude in Section \ref{sec:conclusions}.

\section{Bayesian framework for associating gravitational wave events with AGN flares}\label{sec:method}

In this section we describe a formalism for confidently associating GW events to AGN flares. The goal is to understand how many observations are needed to confirm the association with high confidence (Bayes factors > 100). The problem can be formulated as a signal versus background problem, where for each GW event from a BBH in an AGN disk we have $N$ expected background flares and $N+1$ total expected flares. 

\subsection{$\lambda$: the fraction of BBH that induce an AGN flare}

We consider a similar formalism to that described in \citet{Morgan_2019} for associating IceCube neutrinos to core-collapse supernovae. In our case, we substitute the IceCube neutrinos with GW events, which may produce signal flare if they come from an AGN, and the background supernovae with background AGN flares. We modify the formalism to be fully Bayesian, deriving a posterior probability distribution for the parameter of interest, and calculating Bayes factors.

Let $\lambda$ be the fraction of GW events that are associated with AGN flares, $\lambda = P(\mathrm{AGN} \mid \mathrm{GW})$. Given a GW event $i$ at location $(\Omega_i^\mathrm{GW}, z_i^\mathrm{GW})$ and merger time $t_i^\mathrm{GW}$, the number density of AGN flares per solid angle $\Omega$ and per redshift $z$ within some time period $t^\mathrm{AGN} - t_i^\mathrm{GW} < T$ is given by:
\begin{equation}
\begin{split}
    \frac{dN_i}{d\Omega dz}\left(\Omega, z \mid \Omega_i^\mathrm{GW}, z_i^\mathrm{GW}, \lambda, T, \frac{dB}{d\Omega dz dt}\right) =\\ 
    =\lambda \delta(\Omega_i^\mathrm{GW}-\Omega) \delta(z_i^\mathrm{GW}-z) +  T\frac{dB}{d\Omega dz dt}(\Omega, z),\label{eq:dNAGN}
\end{split}
\end{equation}
where $\delta$ is the Dirac delta function. In other words, the distribution of AGN flares $\frac{dN_i}{d\Omega dz}$ can be modeled as a mixture between an AGN flare at the same position as the GW event (expected number $0 \leq \lambda \leq 1$) and the background number density of AGN flares within a time period $T$, $T\frac{dB}{d\Omega dz dt}$. Here, $\frac{dN_i}{d\Omega dz}$ refers to the \emph{astrophysical} (in other words, intrinsic) distribution of AGN flares, rather than the \emph{observed} distribution. These differ by a factor of the detection efficiency, $P_\mathrm{det}^\mathrm{AGN}(\Omega, z)$. More generally, we may consider the luminosity distribution together with the spatial density of flares in Eq.~\ref{eq:dNAGN}, modeling $dN / d\Omega dz dL$, and $P_\mathrm{det}^\mathrm{AGN}$ may depend on the apparent magnitude corresponding to $L$ and $z$. Note that the luminosity of the signal AGN flare may depend on properties of the BBH, e.g. the total mass $M_\mathrm{tot}$, in which case this can be incorporated into the model of Eq.~\ref{eq:dNAGN}.

For a given GW event with data $x_i^\mathrm{GW}$, the sky location and redshift are imperfectly measured with some joint posterior probability distribution $p(\Omega_i^\mathrm{GW}, z_i^\mathrm{GW} \mid x_i^\mathrm{GW}) \propto p(x_i^\mathrm{GW} \mid \Omega_i^\mathrm{GW}, z_i^\mathrm{GW}) p(\Omega_i^\mathrm{GW}, z_i^\mathrm{GW})$. Realistically, we only consider the density of AGN flares within some volume around the GW event (e.g. the 90\% volume of $p(\Omega_i^\mathrm{GW}, z_i^\mathrm{GW} \mid x_i^\mathrm{GW})$) and accordingly normalize the background number density within this volume. We assume the location of the AGN is perfectly measured. The joint likelihood of observing the GW data $x_i^\mathrm{GW}$ and $k$ AGN flares with positions $\left\{\Omega_{ij}^\mathrm{AGN}, z_{ij}^\mathrm{AGN}\right\}_{j=1}^{k}$, marginalizing over the uncertain position of the GW source $(\Omega_i^\mathrm{GW}, z_i^\mathrm{GW})$, is given by an inhomogeneous Poisson process:
\begin{align}
\label{eq:likelihood}
    \mathcal{L}_i & \equiv p\left(\left\{\Omega_{ij}^\mathrm{AGN}, z_{ij}^\mathrm{AGN}\right\}_{j=1}^{k}, x_i^\mathrm{GW} \mid \lambda, R_B \right) \\
    &= \int p\left(\left\{\Omega_{ij}^\mathrm{AGN}, z_{ij}^\mathrm{AGN}\right\}_{j=1}^{k}, \Omega_i^\mathrm{GW}, z_i^\mathrm{GW}, x_i^\mathrm{GW} \mid \lambda, R_B  \right) d\Omega_i^\mathrm{GW} d z_i^\mathrm{GW} \\ 
    &= \prod_{j = 1}^k \Bigg[ \int p(x_i^\mathrm{GW} \mid \Omega_i^\mathrm{GW}, z_i^\mathrm{GW})p_0(\Omega_i^\mathrm{GW}, z_i^\mathrm{GW}) \times \\
    & \qquad \frac{dN_i}{d\Omega dz}\left(\Omega_j, z_j \mid \Omega_i^\mathrm{GW}, z_i^\mathrm{GW}, \lambda, R_B \right)  d\Omega_i^\mathrm{GW} dz_i^\mathrm{GW}\Bigg] e^{-\mu_i}  \, ,
\end{align}
 where $p_0(z, \Omega)$ refers to the prior on the redshift and sky position of the GW source, we define the background rate $R_B=T \frac{dB}{d\Omega dz dt}$ for simplicity of notation, and $\mu_i$ is defined below. The background term does not carry the GW term because it does not depend the GW position and distance, so that the GW part integrates to 1 in the marginalization over $\Omega_i^\mathrm{GW}, z_i^\mathrm{GW}$.
In the above, $\mu_i$ refers to the expected number of \emph{observed} AGN flares: 
\begin{equation}
    \mu_i \equiv \int \frac{dN_i}{d\Omega dz} P_\mathrm{det}^\mathrm{AGN}(\Omega,z) d\Omega dz\, .
\end{equation}
The background term in $\frac{dN_i}{d\Omega dz}$ does not depend on $\lambda$, so if we are interested in the posterior over $\lambda$, we can consider only the first term in the right-hand-side of Eq.~\ref{eq:dNAGN} when computing $\mu_i$.
Finally, the likelihood becomes:
\begin{widetext}
\begin{equation}
\mathcal{L}_i 
    \propto \prod_{j = 1}^k \left[ \lambda p(x_i^\mathrm{GW} \mid \Omega_{ij}^\mathrm{AGN}, z_{ij}^\mathrm{AGN})p_0(\Omega_{ij}^\mathrm{AGN}, z_{ij}^\mathrm{AGN}) + R_B\left(\Omega_{ij}^\mathrm{AGN}, z_{ij}^\mathrm{AGN} \right)\right] e^{-\mu_i}    
\end{equation}
\end{widetext}
In the cases where no AGN flare is detected in a follow-up (at a location at which the GW localization likelihood has nonzero support), the likelihood of that specific follow-up reduces to:
\begin{equation}
    \mathcal{L}_i \propto e^{-\mu_i},\label{eq:noflare}
\end{equation}
which tends to prefer lower values of $\lambda$, and it is therefore also informative to perform a follow-up that does not detect any flares.
Note that the fraction of GW events with associated AGN flares, $\lambda$, and the number density of background AGN flares, $\frac{dB}{d\Omega dz dt}$ are common to all GW events $i$. For example, we can measure the posterior probability on $\lambda$ by combining observations from $N$ GW events:
\begin{equation}
\begin{split}
    p\Bigg(\lambda \mid & \left\{x_i^\mathrm{GW}\right\}_{i = 1}^N, \left\{ \left\{\Omega_{ij}^\mathrm{AGN}, z_{ij}^\mathrm{AGN}\right\}_{j=1}^{k} \right\}_{i = 1}^{N}, T, \frac{dB}{d\Omega dz dt}\Bigg) \\ 
    & \propto p(\lambda) \prod_{i = 1}^N \mathcal{L}_i.
\end{split}
\end{equation}
With enough GW events, we will be able to measure $\lambda$ and confidently determine whether $\lambda > 0$; in other words, whether a non--zero fraction of GW events are associated with AGN flares.

For a specific GW event $i$ with AGN counterpart $ij$, the probability $p^{\rm GW-AGN}_{ij}$ that the AGN flare is associated with the GW event is given by:
\begin{equation}
\label{eq:odds}
   p^{\rm GW-AGN}_{ij} = \frac{\lambda p(\Omega_{ij}^\mathrm{AGN}, z_{ij}^\mathrm{AGN} \mid d_i^\mathrm{GW})}{\lambda p(\Omega_{ij}^\mathrm{AGN}, z_{ij}^\mathrm{AGN} \mid d_i^\mathrm{GW}) + T\frac{dB}{d\Omega dz dt}\left(\Omega_{ij}^\mathrm{AGN}, z_{ij}^\mathrm{AGN} \right)}.
\end{equation}
This can be inferred jointly with $\lambda$.

In the above, when writing $p(\Omega, z \mid x^\mathrm{GW})$, we have assumed perfect knowledge of the cosmological parameters $\Vec{\theta}\equiv(H_0,\Omega_m, ...)$. The GW data yield a measurement of the luminosity distance $d_L$, related to $z$ via $\Vec{\theta}$. If we assume a prior distribution $p(\Vec{\theta})$, we must marginalize out this prior:
\begin{align}
    p(\Omega, z \mid x^\mathrm{GW})
    &= \int p(\Omega, z \mid x^\mathrm{GW}, \Vec{\theta})p(\Vec{\theta}) d \Vec{\theta} \\
    &= \int p(\Omega, d_L(z, \Vec{\theta}) \mid x^\mathrm{GW}) p(\Vec{\theta}) d\Vec{\theta}. 
\end{align}
Because an uncertain cosmology implies a larger localization volume for a given GW event, we must ensure that the background rate density is normalized over this larger volume as well, especially if we are using the results to infer cosmological parameters. This will tend to increase the expected number of background AGN flares. 

\subsection{Standard sirens}

For GW events with a counterpart, a unique host galaxy, and therefore a cosmological redshift, can be identified. Events without a counterpart require a marginalization over all potential host galaxies and therefore provide a less precise estimate of cosmological parameters (e.g. \citealt{chen17}). In the case of AGN flares, given the possible contamination of background events, the cosmological parameter estimation problem becomes intermediate between the dark-siren and unique-counterpart cases.

Let us consider $N$ GW events $\left\{ x_i^\mathrm{GW}\right\}_{i = 1}^N$, which have been followed-up with observations of the AGNs in the relevant volumes. Considering the follow-up data $\left\{x_i^\mathrm{AGN}\right\}_{i = 1}^N \equiv \left\{\left\{\Omega_{ij}^\mathrm{AGN}, z_{ij}^\mathrm{AGN}\right\}_{j=1}^{k}\right\}_{i =1}^N$, the posterior on the cosmological parameters $\cosvar$ is:
\begin{align}
    p(\cosvar|\left\{x_i^\mathrm{AGN}\right\}_{i = 1}^N, \left\{x^\mathrm{GW}_i\right\}_{i = 1}^N) & \propto p(\cosvar) \int d\lambda p(\lambda) \prod_i^N \mathcal{L}_i\left(\cosvar, \lambda \right)
\end{align}
Going back to Eq. (\ref{eq:likelihood}) and modifying the likelihood to be conditioned on the cosmology, we get: 
\begin{widetext}
\begin{align}
\label{eq:cosmo-likelihood-single}
    \mathcal{L}_i\left(\cosvar, \lambda \right) \propto \prod_{j = 1}^k \left[ \lambda p(x_i^\mathrm{GW} \mid \Omega_{ij}^\mathrm{AGN}, d_L(z_{ij}^\mathrm{AGN}, \cosvar))p_0(\Omega_{ij}^\mathrm{AGN}, z_{ij}^\mathrm{AGN}) + R_B\left(\Omega_{ij}^\mathrm{AGN}, z_{ij}^\mathrm{AGN}, \cosvar \right)\right] e^{-\mu_i}
     \end{align}
\end{widetext}
If no flares are identified in a follow-up, the likelihood is that of Eq.~(\ref{eq:noflare}). If there is no GW follow-up, the AGN likelihood is uninformative. However, we note that one can substitute the GW prior $p_0(\Omega, z)$ for a galaxy catalog (or equivalently, replace $p_0$ with a galaxy catalog posterior $p (\Omega, z \mid x_\mathrm{gal})$). In this case, the likelihood will reduce to that of the statistical standard siren method~\citep{2012PhRvD..86d3011D,fishbach,Palmese:2020aof}.
Here we have ignored GW selection effects, which play an important role especially for cosmological measurements. To account for GW selection effects,  Eq.~\ref{eq:cosmo-likelihood-single} must be divided by a term $\beta(\cosvar)$, so that it integrates to unity over detectable GW datasets (e.g.~\citealt{mandel}). 

\section{The case of GW190521}\label{sec:190521}

The AGN J124942.3+344929 is not particularly well placed in the LIGO-Virgo GW190521 sky localization map. Nonetheless, the position on the sky has support in line of sight probability. We wish to estimate a probability of chance occurrence.

\begin{table}\centering
\begin{tabular}{ll}
\hline
Event & Volume [Gpc$^3$]\\ % %& 
\hline
GW190814  & $9.2\times 10^{-5}$ \\
GW170814 &  $1.5\times 10^{-4}$ \\
GW190701\_203306 & 0.087 \\
GW190521 & 9.1\\
\end{tabular}\caption{Comoving volume (99\% CI) for the sample of LIGO/Virgo GW binary black hole events considered in this work. Notice that GW190521 has a volume orders of magnitude larger than the other events.}\label{tab:vol}
\end{table}

As the AGN luminosity function is known  over the range of redshifts of interest from, e.g., \citet{Hopkins_2007} \& \citet{Shen2020}, our program is straightforward. For a given search limiting magnitude, integrate down the luminosity function to the luminosity corresponding to that $z$'s magnitude limit, then multiply by the spatial volume of the search area. This yields the average number of quasars in the volume. The statistics of quasar variability may then be assessed to estimate the number of quasars varying over the timescale of interest and the magnitude difference required to be labeled a flare. Simply, the expected number of flares is  $N_{f} = \mathrm{Vol}\cdot \phi \cdot f_{f}$, where $\phi$ is the volume density of quasars and $f_{f}$ is the fraction of those that vary enough to be labeled a flare.

GW190521 has a spectacularly large localization volume, due to the large 90\% sky localization of 936 deg$^2$ and the large mean luminosity distance of $3.92^{+2.19}_{-1.95}$ Gpc \citep{gwtc2}. We calculate the spatial volume of localization of GW190521 using the the sky map from \citet{GW190521} and the software from \citep{bayestar,Singer_2016,Singer_supp}. We find that the 90\% credible interval (CI) comoving volume is 4.1 Gpc$^3$, and the 99\% volume is 9.1 Gpc$^3$. For context, see the other localization volumes in Table~1. For our $N_{f}$ calculation we take $\mathrm{Vol} = 4.1$ Gpc$^3$.

\begin{figure}
\centering
\includegraphics[width=1\linewidth]{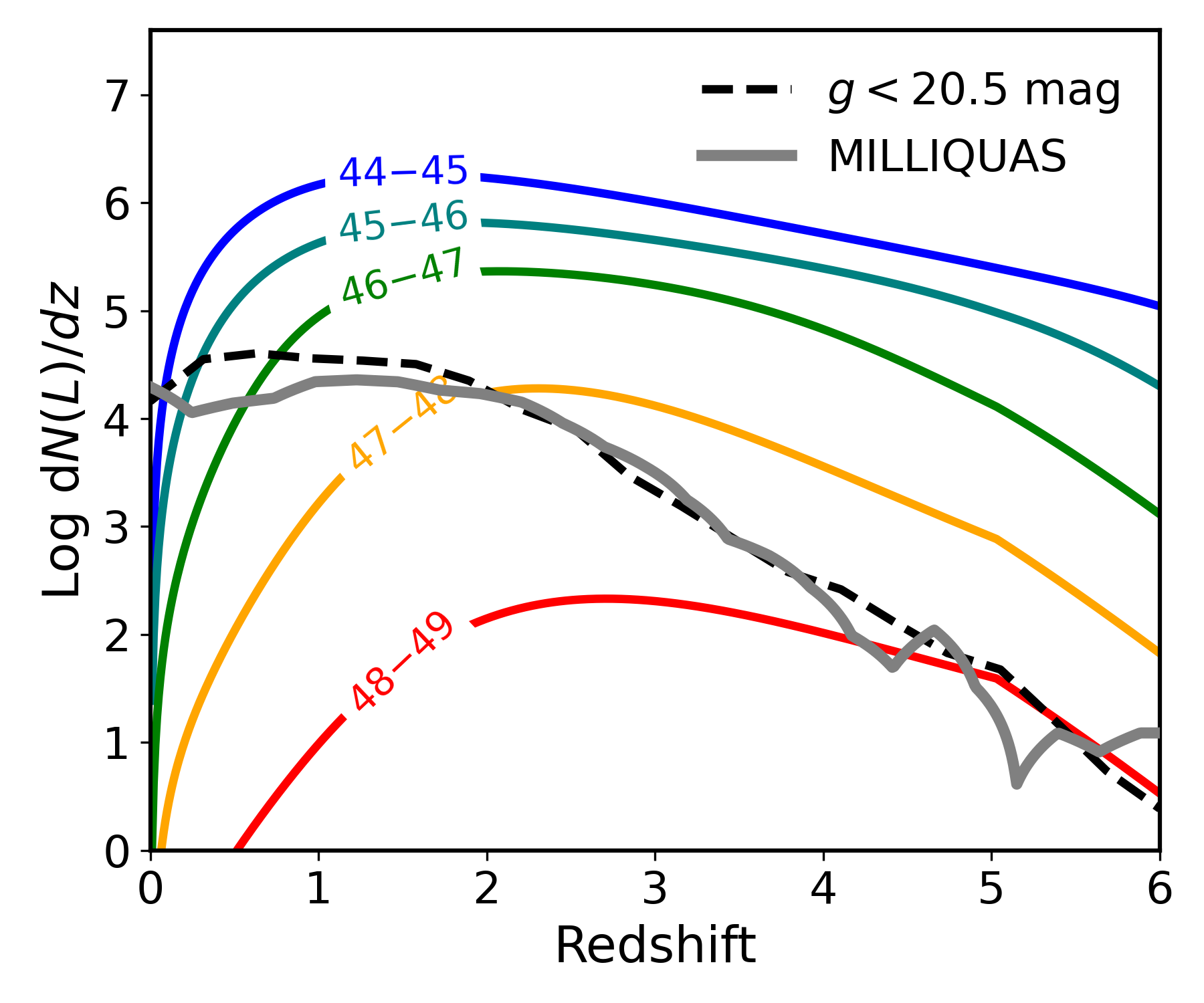}
\caption{Redshift distribution of the number of AGNs using the quasar luminosity function from \citet{Hopkins_2007} in an area of the sky corresponding to the 90\% CI GW190521 sky map. The thick solid lines are the redshift distributions in various luminosity intervals (in Log($L_{\rm{bol}}/$[erg s$^{-1}$])). The dashed black line is the redshift distribution with a flux limit of $g<20.5$ AB mag. The grey line is the redshift distribution of AGNs from the MILLIQUAS catalog that lie within the 90\% CI GW sky map. It is clear that there order of $\sim10^4$ AGNs around the redshift $z= 0.64^{+0.28}_{-0.28}$ of the event.}\label{fig:qlf}
\end{figure}

To estimate the number of AGNs in the localization volume, we use the quasar luminosity function (QLF) from \citet{Hopkins_2007}, given in a dual power law form:
$$\phi(L) = \frac{\phi_\star}{(L/L_\star)^{\gamma_1} + (L/L_\star)^{\gamma_2}},
$$ where at $z=1$ the parameters of the fit are $\log\phi_\star = -4.6$ quasars/Mpc$^3/\log(L)$, $\log (L_\star/L_\odot )= 12.6$ , $\gamma_1=-0.4$, $\gamma_2=-2.2$, and where $L$ is the bolometric luminosity. The parameter values vary as a function of $z$ as the quasar population evolves. Note the faint quasar power law is shallow while the luminous quasar power law is steep. In figure~\ref{fig:qlf} we show the 
$dN(L)/dz$ in the 90\% sky area, assuming a concordance cosmology for the volume. The numbers are dominated by quasars at or below the break in the dual power-law. How far down the LF one sees at a given $z$ is an observational question. For the ZTF limiting magnitude of $g<20.5$ AB mag, we integrate down the LF to the corresponding limiting luminosity, assuming $L_{\rm{bol}}=10\ L_{\rm{band}}$, where $L_{\rm{band}}$ is the luminosity in an optical band following \citep{Hopkins_2007}. The result is the black dashed line in Figure~\ref{fig:qlf}, which shows  the  $dN/dz$ of quasars in the area. 
 We find that there are $\sim 34,000$ AGNs in the 90\% localization area out to $z< 1$. For comparison, we show the redshift distribution of quasars in the Million Quasars (MILLIQUAS) Catalog \citep{Flesch2019} in the 90\% localization area. The dominant source of AGN in MILLIQUAS are SDSS quasars. 
 While we have computed the number of AGNs using the luminosity function, its redshift evolution, and an apparent magnitude limit, our numbers are equivalent to considering a uniform AGN number density of $n_{\rm AGN}= 10^{-5}$ Mpc$^{-3}$ (which is lower than the fiducial value of $z\approx 0.2$ type-I AGN considered in B17,  $n_{\rm AGN}= 10^{-4.75}$ Mpc$^{-3}$), %\citealt{Bartos}), 
 since that would translate into a total number of AGNs in the 90\% volume Vol$=4\times10^{9}$ Mpc$^3$ of 40,000. 
 
{We use the prescription of \citet{Hopkins_2007} to calculate the fraction of 
 type-I quasars, now known as optically unobscured AGNs. Originally the difference between type-I and type-II AGN was whether they showed broad+narrow lines (type-I) or only narrow lines in the optical spectrum. This is important when computing the probability of chance coincidence for optical flares, because a flare in the accretion disk is expected to be obscured from view, at least in the optical. It does not necessarily mean that a BBH merger could not happen in an obscured AGN, or that a flare could not be observed as a ``reprocessed'' flare at other, perhaps longer, wavelengths (see \citealt{Kool_2020} for an example of transient candidate in an obscured AGN). We find that there are $\sim 7,400$ Type I AGNs in the 936 deg$^2$ down to $g<20.5$. For our $N_{f}$ calculation we take $\mathrm{Vol}\cdot \phi = 7,400$ quasars. This is different from the $\sim 3,000$ AGNs considered in G20 for two main reasons. First, they consider the volume covered by ZTF, which is roughly half of the total volume in the preliminary sky map \citep{GCN2}. Secondly, we use the updated sky map from LIGO/Virgo, which encompasses a larger volume than the LALInference map used in G20. 
 Alternatively, if one wants to take into account AGNs below the $g<20.5$ limit, and consider all Type I AGNs down to our bolometric luminosity limit of $10^{44}$ erg s$^{-1}$, the number density is $\sim 10^{-4.5}$ Mpc$^{-3}$, which results in a number of AGNs in the 4 Gpc volume of $\mathrm{Vol}\cdot \phi \sim 130,000$. }

\begin{figure}
\centering
\includegraphics[width=1\linewidth]{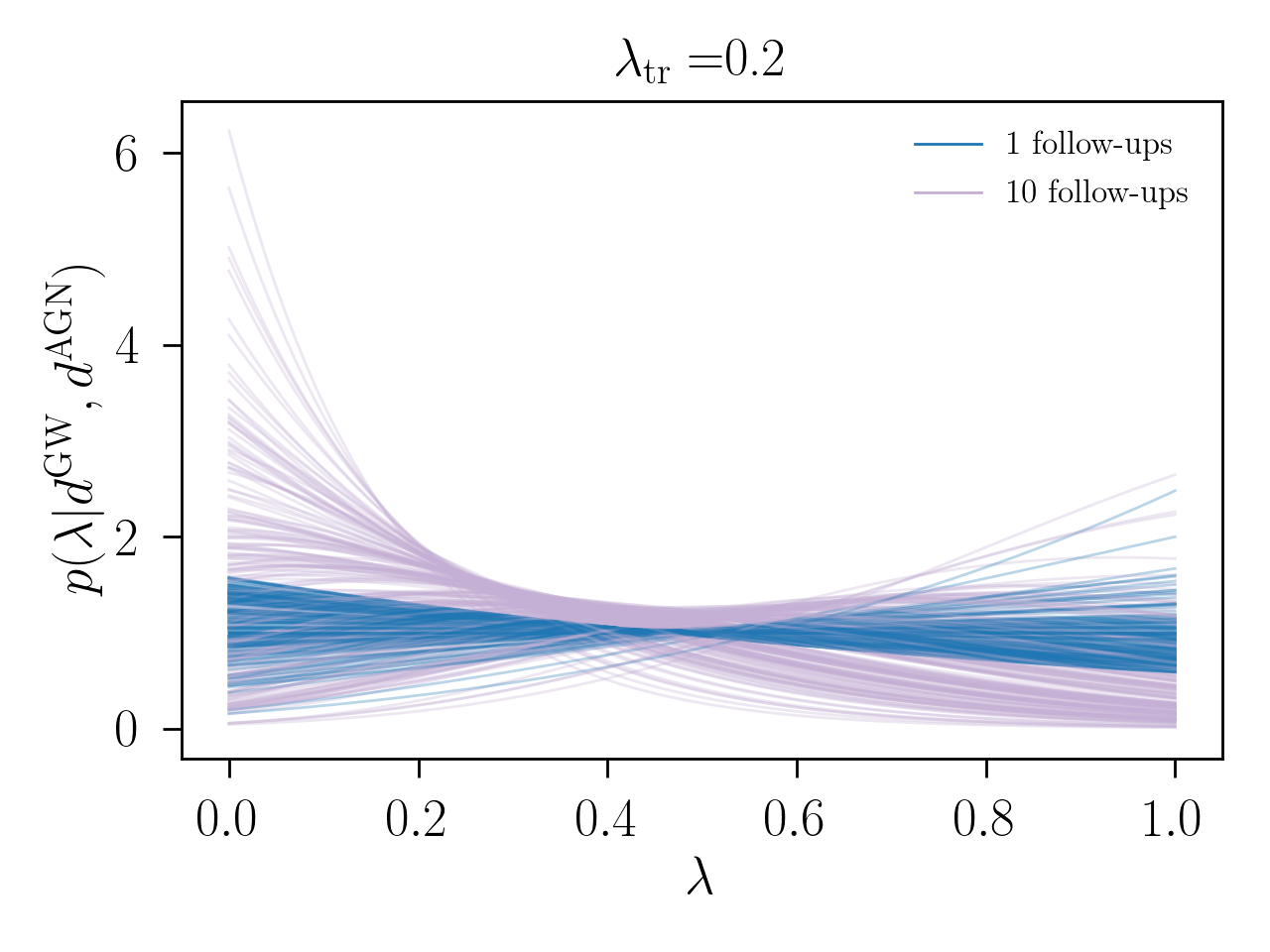}
\includegraphics[width=1\linewidth]{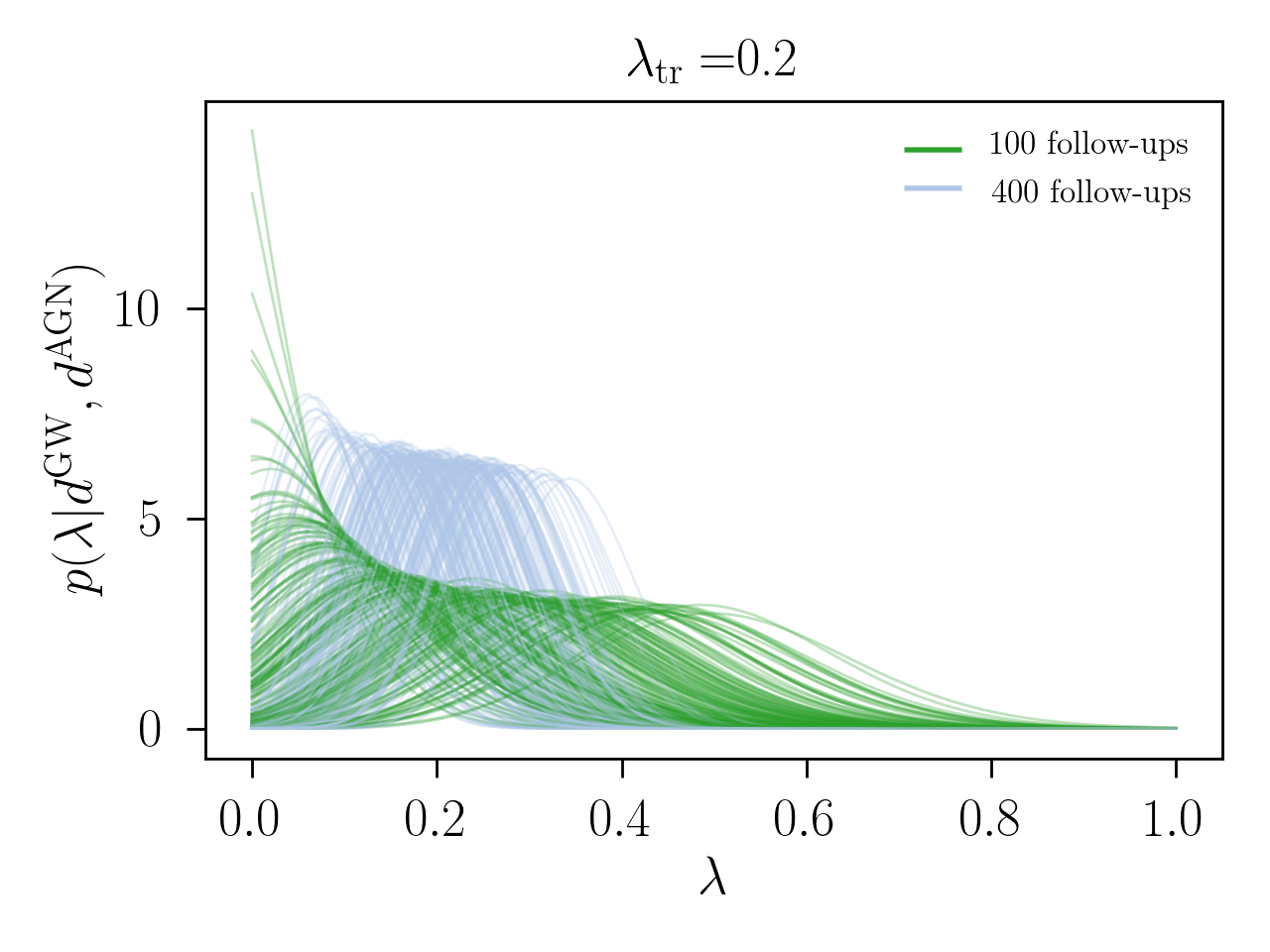}
\caption{Posteriors for $\lambda$ from 200 sets of simulations of 400 follow-up observations of GW190521--like events for an input value of $\lambda=0.2$, as an example of the method. The posteriors are derived for 4 values of number of follow-ups, and for a uniform prior in $\lambda$ between 0 and 1. It is clear that the posterior becomes more constraining around the true value of $\lambda$ as the number of follow-up observations increases.}\label{fig:post}
\end{figure}

The labelling of a quasar variability event a flare is a judgement. Most or all quasars vary; searching for point sources that vary is one of the very best ways to find quasars. A common model for quasar variability is the damped random walk (DRW), yet this is a particular model and questions about its general applicability remain in the literature - see e.g. \cite{Kasliwal2015} for the question of short timescales, and  \citealt{kozowski} for a measurement of the distribution of quasar variability  power law indices about and biased from the DRW index. \citet{Graham} use a DRW model to estimate the probability of chance occurrence, and the literature suggests treating this with caution.   We will instead use structure functions (SF), which are a more general description of variability; see \citealt{kozowski} for a review.  The structure function is
\begin{equation}
    \mathrm{SF}(\Delta t) = \mathrm{SF}_0 \left(\frac{\Delta t}{\Delta t_0}\right)^\gamma,
\end{equation} 
where $\mathrm{SF}$ is measured in magnitudes ($\Delta m$), $\mathrm{SF}_0$ is the $\Delta m$ measured at some time $\Delta t_0$, say 100 days, and $\gamma$ is the power law index.
The SF is not a physical model, but an observational, statistical  description of AGN variability. The DRW is a special case of the SF:
\begin{equation}
\mathrm{SF}(\Delta t) = \mathrm{SF}_\infty \left(1 - e^{-\Delta t/\tau}\right)^{0.5},\label{eq:DRW}
\end{equation}
where $\mathrm{SF}_\infty$ is measured as some time suitably long compared to the problem. The timescale $\tau$ may be the timescale of a model related, for example, to black hole mass. Of note is that the DRW the power law index is fixed at 0.5; the SF  measures this as its $\gamma$ parameter.

\citet{Graham} report 
AGN J124942.3+344929 varied by $\approx 0.4$ mag over 50 days.
We estimate the probability of this using SF measurements from %\citet{Caplar_2017} and \citet{Kimura_2020}, who use PTF and HSC optical data respectively. 
\citet{Kimura_2020}, who present HSC optical data for a robust sample of AGNs down to $r\lesssim 23.5$. 
Their Figure 18 shows the SF versus $\Delta t$. For $g$ band and $\Delta t_0\sim$ 30 days, they find SF$\approx0.15$. 
We will use $\Delta m = 0.4$ mags. One interprets the SF as the timescale dependent standard deviation $\sigma$ of a normal distribution centered on 0, which describes the probability distribution of having a $\Delta m$ change in magnitude for an AGN, and one calculates the onesided probability corresponding to $\Delta m>0.4$,  $f_f \sim 10^{-4}$.  By comparison, \citet{Graham} estimated that the chance of their flare model fitting any ZTF AGN lightcurve is $\sim 5\times 10^{-6}$. The difference between the flare probability of \citet{Graham} and ours is the fact that they required a fit with a specific flare shape to the available sample of AGNs, while we only require a magnitude change over a timescale.

We are now ready to compute  $N_{f} = \mathrm{Vol}\cdot \phi \cdot f_{f}$. Our estimate is $N_f = 7,400 \times10^{-4}\sim 0.74$ flares in the area. 
This translates in a 70\% probability of chance coincidence of an AGN flare of $\Delta m >0.4$ in the 90\% GW localization. Even using the 
\citet{Graham} estimate that the chance of their flare model fitting any ZTF AGN lightcurve is $\sim 5\times 10^{-6}$ using our calculated unobscured AGN numbers, this implies a  $\sim 4\%$ probability of chance coincidence. {To further understand if the flare probability we find is reasonable for this specific AGN, and for comparison with G20, we use the long-term available data to fit a DRW model. We fit the SF to the unbinned CRTS \citep{2011arXiv1102.5004D} and ZTF \citep{Masci_2018,Bellm_2018} data following the method of \citet{Kelly2009} using the \textsc{celerite} code \citep{Foreman-Mackey2017}. We mask out the light curve portion associated to the flare in 2019, and find that the parameters of Eq. (\ref{eq:DRW}) are $\ln ({\rm SF}_{\infty}/{\rm mag})=-0.75^{+0.16}_{-0.10}$ and $\ln (\tau/{\rm days})=5.91^{+0.85}_{-0.58}$. For a timescale of 30 days, the maximum likelihood values of the DRW correspond to a SF of $\sim 0.13$, thus the probability of observing a flare of magnitude 0.4 in this AGN is $\sim 10^{-3}$. We conclude that the observed flare could be associated with stochastic AGN variability, and that our $f_f$ calculation for a generic AGN is reasonable also for the AGN in question.}

Clearly how one does the $f_f$ calculation matters, but we argue that probabilities of $4\%$ to $70\%$ of a chance occurrence suggests that the flare in AGN J124942.3+344929 is consistent with being a background flare. It is worth recalling that ZTF did not cover the $90\%$ spatial localization, as their observations covered $\approx 50\%$ of the probability in sky localization.

An important note is that our simple calculation is conservative, in the sense that a larger number of AGNs could be considered. We did 
 not include low-luminosity AGNs $L_{\rm{bol}}<10^{44}$ erg s$^{-1}$, which are more abundant than quasars in particular at low redshift (e.g. \citealt{2005hao}). {In addition, we have considered a minimum $\Delta m=0.4$, while a lower cut, say at 0.3, would result in an order of magnitude more probable flares, and therefore an order of magnitude more expected flares, bringing the probability of chance occurrence in the GW region to 40-100\%.}

An extension to our analysis is to use the spatial distribution of GW distance and distance uncertainty in the calculation of the limiting luminosity of the AGNs observed from the AGN luminosity function, and to do so over the (much larger) 99\% confidence level localization. We do not expect this to significantly affect the result, and the high probability of chance coincidence does not provide sufficient motivation to pursue it. However, it is worth pursuing the question of how to constrain the possibility that AGN accretion disks do provide the site of BBH mergers.

\section{Constraining the fraction of BBH inducing an AGN flare}\label{sec:future}

The question of which formation channel(s) are responsible for creating BBH systems is hotly debated. It is of considerable interest to evaluate the fraction of BBH events that come from AGN disks. In this section we show the results of applying the Bayesian method described in Section \ref{sec:method} to simulations of future GW events in pursuit of the number of events necessary to constrain the fraction $\lambda$ of BBH events that produce AGN flares.

First, we assume GW events like GW190521. In this case, we draw the distances and sky positions of the signal events from the sky map posterior samples of GW190521. The background events are drawn from a Poisson distribution with an expectation value of $4 \times10^9 \times 10^{-4}\times 10^{-4.5}$ following the number of flares from the HSC $g-$band SF and the number of AGNs in the 90\% volume. The choice of the number of AGNs that would contribute to the background flares depends on a number of factors, including the depth of the survey, the wavelengths observed, and the redshift of the event. We therefore decide in the following to make the most conservative assumption, and assume that the average number density of those AGNs is $10^{-4.5}$ Mpc$^{-3}$, which would include all Type I AGNs down to $L_{\rm bol}=10^{44}$ erg s$^{-1}$.
We generate 200 sets of simulations of up to 800 follow-ups for different input values of $\lambda$ between 0 and 1. As an example, the $\lambda$ posteriors for a truth value of $\lambda_{\rm tr}=0.2$, for 4 values of number of follow-ups, and for a uniform prior in $\lambda$ between 0 and 1 are shown in Fig. \ref{fig:post}. It is clear that the posterior becomes more constrained around the true value of $\lambda$ as the number of follow-ups increases.

\begin{figure}
\centering
\includegraphics[width=1\linewidth]{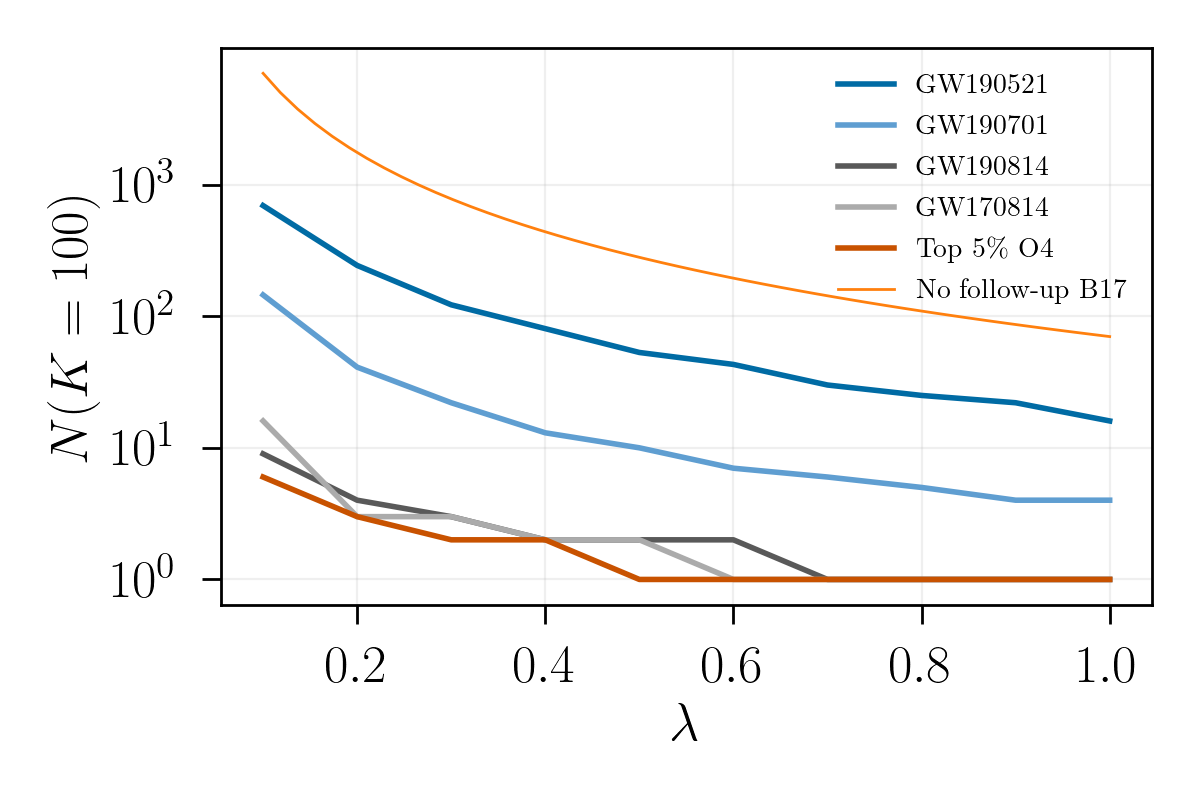}
\caption{Number of GW events to be followed-up to obtain a confident association (i.e. for a Bayes factor $K=100$) between AGN flares and BBH events, as a function of the true input value of $\lambda$ in the simulations. Each curve corresponds to a different fixed GW sky map, with a range of localization volumes (from $\sim10^{-4}$ to 10 Gpc$^3$ at 99\% CI). This figure applies to the ideal case in which the AGN flares associated to the GW events are all detectable (we assume a magnitude limit $g<20.5$). The orange line shows a comparison with the result from \citet{Bartos} (B17), who do not consider follow-up observations of AGN flares, but instead compare the GW localizations with AGN positions. The other differences with B17 are that their $N$ corresponds to the number of events needed to reach 3$\sigma$ using a $p-$value statistical method, and that they consider a slightly lower AGN number density. }\label{fig:k}
\end{figure}

We repeat the same procedure using better localized sky maps, namely those of GW190701\_203306 (\citealt{gwtc2}; 99\% CI comoving volume 0.087 Gpc$^3$), GW170814 (\citealt{gw170814}; 99\% CI volume $1.5\times 10^{-4}$ Gpc$^3$), and GW190814 (\citealt{GW190814}; 99\% CI volume $9.2\times 10^{-5}$ Gpc$^3$). We scale the expected number of background events based on each event's volume. We then compute the Bayes factor $K$ for a model with $\lambda>0$ (which is taken to be the mode $\bar{\lambda}$ of the $\lambda$ posterior) versus a model with $\lambda=0$ using the likelihoods obtained from our simulations, as the Savage-Dickey density ratio:
\begin{equation}
    K = \frac{p(x|\lambda=\bar{\lambda})}{p(x|\lambda=0)}\, ,
\end{equation}
where $x$ is the GW and AGN data.
We consider the AGN association (i.e. $\lambda>0$, because at least a fraction of the BBH come from AGNs) to be confident if $K>100$.
The number of follow-up observations needed to reach this requirement as a function of the true value of $\lambda$ given in input is shown in Fig. \ref{fig:k} for different sky maps. It is clear that for poorly localized events like GW190521, tens of GW follow up campaigns are required in order to make a confident association, even in the most optimistic case where $\lambda_{tr}=1$ and we can detect all AGNs where the BBH merger could happen. Only a few to tens of events are needed to make a confident association for better-localized events such as GW170814 or GW190814 down to $\lambda_{tr}=0.1$. This is an obvious consequence of the fact that the number of background contaminants scales with the comoving volume, and that the localization volume of these events is orders of magnitude lower than the one of GW190521.

We compare our results with the predictions by B17 who uses only GW localizations (i.e. without follow-up observations) to probe the origin of BBH mergers and to potentially associate them with AGNs. Fig. \ref{fig:k} shows that we find a similar scaling relation of the number of events required to reach $K=100$ as a function of $\lambda$ as theirs, although slightly less steep than $N (K=100) \propto \lambda^{-2}$.
The main differences with the B17 predictions, other than follow-up observations, are that their $N$ corresponds to the number of events needed to reach 3$\sigma$ using a $p-$value statistical method (so that here the scaling of the number of events needed could be different as we are not directly considering the width of a distribution), and that they consider a fixed (in redshift) and lower AGN number density.

Next, we consider the expected constraints that can be derived from the upcoming LIGO/Virgo/KAGRA observing run O4, expected to start in 2022. We consider the AGN flare observable $\Delta m$, the change in the AGN magnitude over time, as above, and use it to compute the expected number of background events using the same SF. We assume that for BBH mergers in AGN disks, the observed $\Delta m$ depends on the total source-frame BBH mass according to the prescription in \citet{McKernan19}, also used in \citet{Graham}, where the counterpart brightness is proportional to $M_{\rm BBH}^2$. The observed flare for a generic BBH in an AGN disk, expressed in terms of the total mass $M_{\rm BBH, 19}$, the potential counterpart flux $F_c$, and AGN flux $F_{\rm AGN}$ for GW190521 will have: 
\begin{equation}
    \Delta m = -2.5 \log \Big[ 1+\frac{M_{\rm BBH}^2}{M_{\rm BBH, 19}^2} \Big( \frac{F_c}{F_{\rm AGN}}\Big)_{\rm GW190521}\Big]\, .
\end{equation}
Ideally, one would rescale the AGN flux as well based on the AGN luminosity function for each potential host AGN in the simulations. However, using the flux of SDSS J124942.3+344929 as a ``fiducial'' AGN flux is reasonable because its bolometric luminosity is $\log (L/L_\odot)\approx 12.3$, and therefore very close to the $L_*$ value from \citet{Hopkins_2007} at $z=0.5$, where $\log (L_*/L_\odot)= 12.24$. This implies that AGNs brighter than SDSS J124942.3+344929 are  rare, while the majority of AGN we consider here will be less bright, and therefore a counterpart of the same luminosity would be even more easily detected.

We simulate GW events using the \texttt{BAYESTAR} software \citep{bayestar,Singer_2016,Singer_supp}, also based on tools from \texttt{LALSuite} \citep{lalsuite}. We assume sensitivity curves for Advanced LIGO and Virgo at O4 sensitivity as published in~\citet{ligoprospects} (\url{https://dcc.ligo.org/LIGO-T2000012/public}). We also consider the addition of KAGRA during O4 with the sensitivity curve from \url{https://dcc.ligo.org/LIGO-T2000012/public}, having a BNS range of $\sim 80 $ Mpc. The simulation includes 10,000 BBH following a distribution that is uniform in comoving volume, assuming a \citet{planck18} cosmology. We assume IMRPhenomD waveforms both for the injections and reconstructions.
We modify the \texttt{BAYESTAR} code so that the primary BHs follow a mass function based on the best fit from the ``power-law + peak" model of \citet{O3a_population}. The primary mass distribution is described by a power-law with index 1.6, plus a Gaussian peak centered on 33 $M_\odot$ and with standard deviation of $6 ~M_\odot$. The events following the power law comprise $90\%$ of the sample, while the events from the Gaussian peak are 10\%. The maximum BH mass considered is $100~M_\odot$, and the BHs follow a uniform spin distribution between $(-1,1)$. After the 10,000 injections are made, we run a matched-filter search to retrieve the detected events. A detection is made when at least 2 detectors reach a single--detector signal--to--noise ratio SNR$>4$ and the network SNR is $>12$. Gaussian noise is added to the measured SNR. In the last step, we reconstruct \texttt{BAYESTAR} skymaps for the detection. The reconstruction is made assuming a distance prior which scales as $\propto d_L^2$, where $d_L$ is the luminosity distance. 

We select BBH with $M_{\rm tot}>50 ~M_\odot$, because they are assumed to give rise to brighter flares than lower mass events. Out of 2401 simulated detections, 1131 events meet the mass cut. For each event, we calculate the expected number of background events based on the 90\% CI volume from the \texttt{BAYESTAR} reconstruction and the SF at the expected flare magnitude given the mass of the binary, assuming a conservative number density of AGNs of $10^{-4.5}$ Mpc$^3$. The final number we use to define the events of interest is the number of expected background events, which is ultimately what defines how quickly one can reach a confident association. 

Using the population fits from~\citet{O3a_population}, we infer that the astrophysical merger rate for BBH systems with $M_{\rm tot}>50~ M_\odot$ is 3--6 Gpc$^{-3}$ yr$^{-1}$ (90\% credible interval). 
At the sensitivity expected for O4, our simulations predict $59$--$117$ detections in this mass range per year of observation. We find that $\sim 4-6$ events per year ($\sim 5\%$ of all $M_{\rm tot}>50 M_\odot$ BBH detections) will be better (i.e. they will have a smaller number of expected background flares) than the forecasts above labeled as GW170814, 7-10 per year ($\sim 8\%$) will be better than the forecasts labeled as GW190701, and $\sim 19-28$ ($\sim 24\%$) will be better than GW190521. It is clear from Fig. \ref{fig:k} that the top 5\% events in the mass range of interest are those that will provide the most significant constraints: following-up these well localized and brighter events will allow us to confidently say if $\lambda>0$, at least in the case where the true value of lambda is $\lambda_{\rm tr}\gtrsim 0.1- 0.2$ (3-6 events required) for a year-long O4 run. On the other hand, it is clear from Fig. \ref{fig:k} that if no follow-up is observations are obtained, the GW-AGN association is likely to only be possible if $\lambda\gtrsim 0.4$ during a one-year-long O4 run, since $>500$ events are needed for $\lambda<0.4$, at least in this fiducial case. The motivation for pushing down to lower $\lambda$ values is that current detections suggest that there are multiple BBH formation channels at play~\citep{O3a_population, 2020arXiv201103564W, zevin2020channel,2021arXiv210212495B}, and that one single channel does not contribute to more than $\sim 70\%$ of all the BBH \citep{zevin2020channel}, i.e. it is likely that $\lambda_{\rm tr}<0.7$.

We note that the method for choosing the top 3\% events for this forecast does not solely rely on the localization volume, but also on the total mass of the binary. We therefore suggest that an estimate of the BBH total mass could be an interesting parameter to share with the astronomical community during the next LIGO/Virgo/KAGRA observing runs.

\begin{figure}
\centering
\includegraphics[width=1\linewidth]{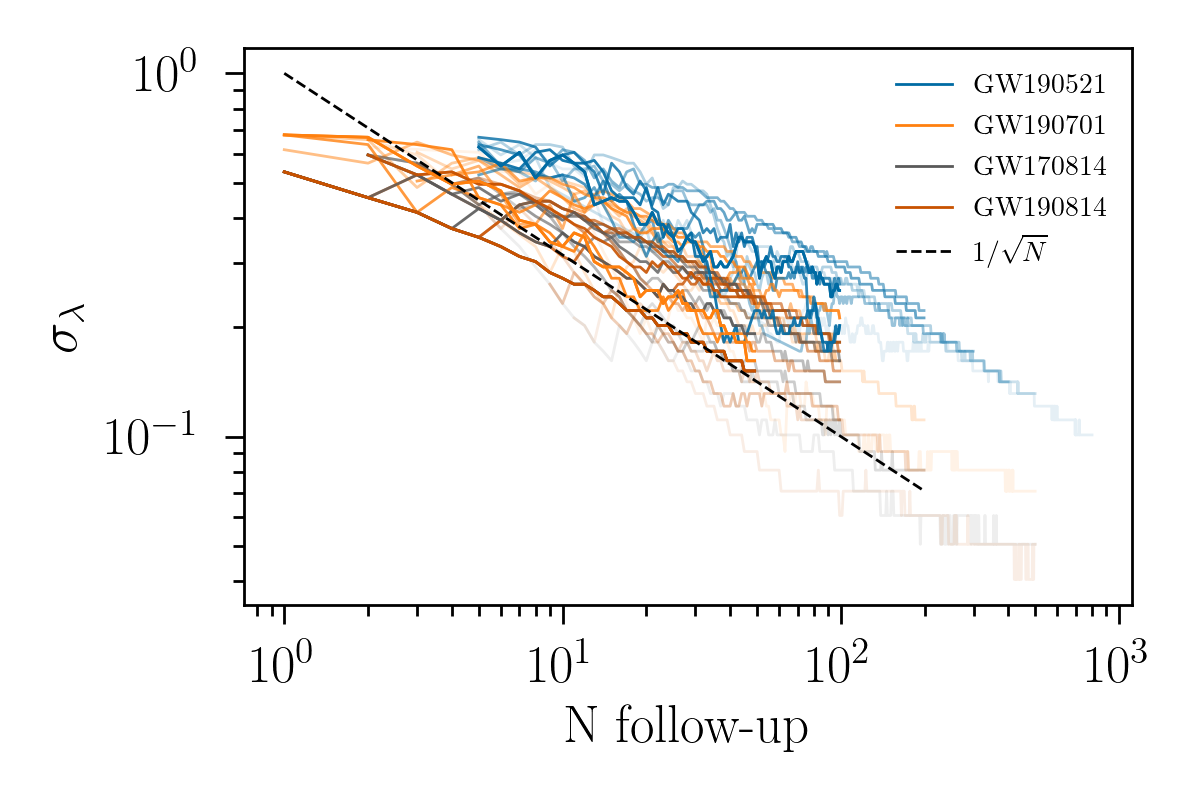}
\includegraphics[width=1\linewidth]{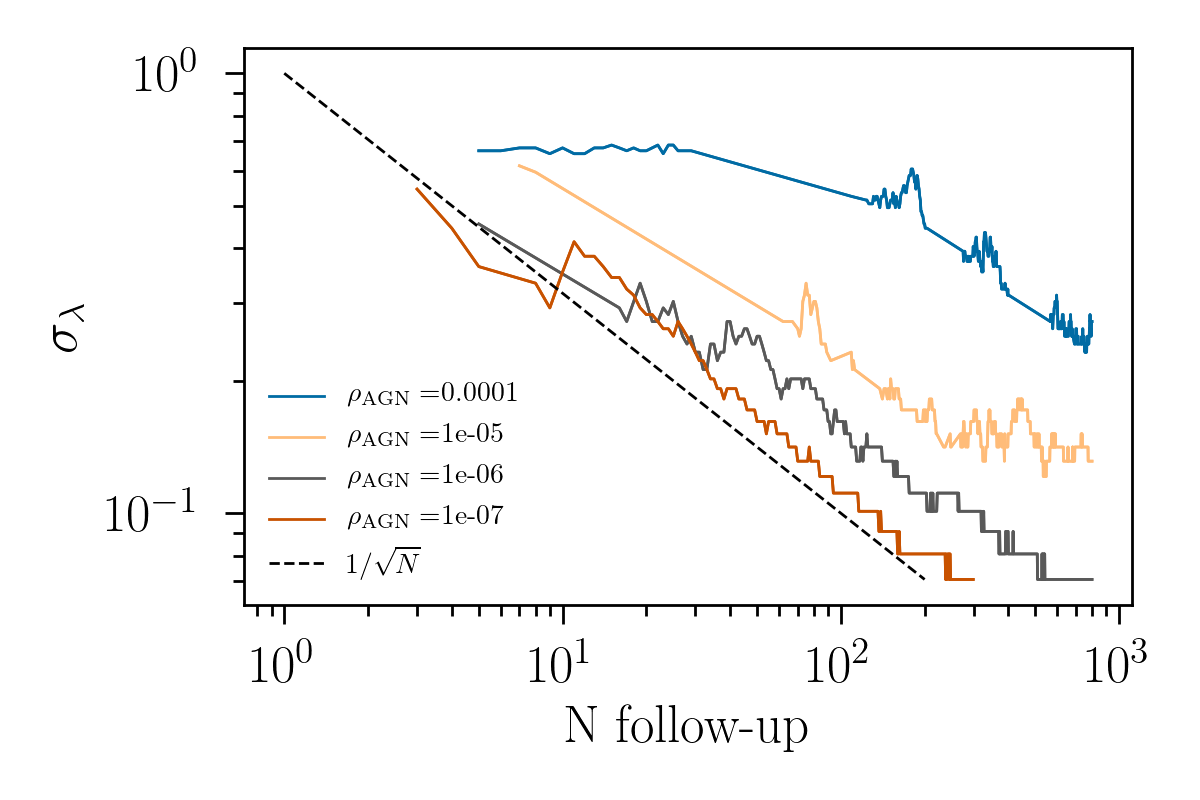}
\caption{\emph{Top:} Expected uncertainty (68\% CI) on the fraction $\lambda$ of GW BBH that can be associated to AGN flares, as a function of number of follow-up observations performed, for sky maps of different GW events. The transparency of the lines is given by the input parameter of $\lambda$ assumed in the simulations, between 0.1 (lighter lines) to 1 (heavier lines). A $1/\sqrt{N}$ scaling is shown for reference. Simulations using higher $\lambda$ values or smaller localization volumes end earlier since they need less events to reach the required precision. \emph{Bottom:} same as the plot above, except the sky map is fixed to that of GW190521, $\lambda$ is fixed to 0.1, and the number density of AGNs is changed. }\label{fig:lamerr}
\end{figure}

In the top panel of Fig. \ref{fig:lamerr} we show the scaling of the uncertainty on $\lambda$ from the simulation as a function of the number of events followed-up, for different events. As expected, better localized events have a smaller number of background flares, and reach a better precision with a smaller number of follow-ups than events like GW190521. The effect of varying the number of background flares can be better seen on the bottom panel of Fig. \ref{fig:lamerr}, where the map is GW190521 for all lines and the input value of $\lambda$ is fixed to 0.1, while the density of AGN (each considered with its own probability of flaring) is changed between 10$^{-7}$ and 10$^{-4}$ Mpc$^{-3}$. The scaling roughly follows $\propto 1/\sqrt{N}$, where $N$ is the number of follow up observations considered. 

It should be noted that the results above are valid when considering bright flares with a probability computed from structure functions (or any other flare happening with a probability of $\sim 10^{-4}$ in an AGN at any given time), and for AGNs brighter than $L_{\rm{bol}}=10^{44}$ erg s$^{-1}$. We are not aware of a theoretical argument that would set a specific threshold for either the AGN luminosity or flare magnitude, so we have showed most of our results based on the follow-up details of GW190521 and this luminosity limit. As more sophisticated theoretical modeling of the BBH merger mechanism is AGNs becomes available, it will be possible to rescale our results based on new thresholds.

\section{Cosmological parameter estimation in a noisy source identification environment}\label{sec:futurecosmo}

Using a contaminated sample of AGN flares as GW counterparts, without accounting for chance coincidences, will recover biased cosmological parameters. 
For $H_0$ measurements, the bias depends both on the value of $\lambda$ (lower values of $\lambda$ will result higher contamination of background flares), and on the detection threshold of AGN flares. If we assume that we can see all AGN flares, then most background flares live at larger distances, giving rise to most likely measurements of $H_0$ that are larger than the true value of the Hubble constant.
In reality, it is likely that we will be more sensitive to the lower-redshift flares from magnitude limited sky surveys, and this would tend to bias $H_0$ low rather than high. For well-localized events (similar to GW170814, for example), the rate of background flares is sufficiently low that the probability of chance coincidence is lowered and there is less risk of biasing cosmological measurements. In general, however, the contribution from background flares must be properly accounted for.

\begin{figure}
\centering
\includegraphics[width=1\linewidth]{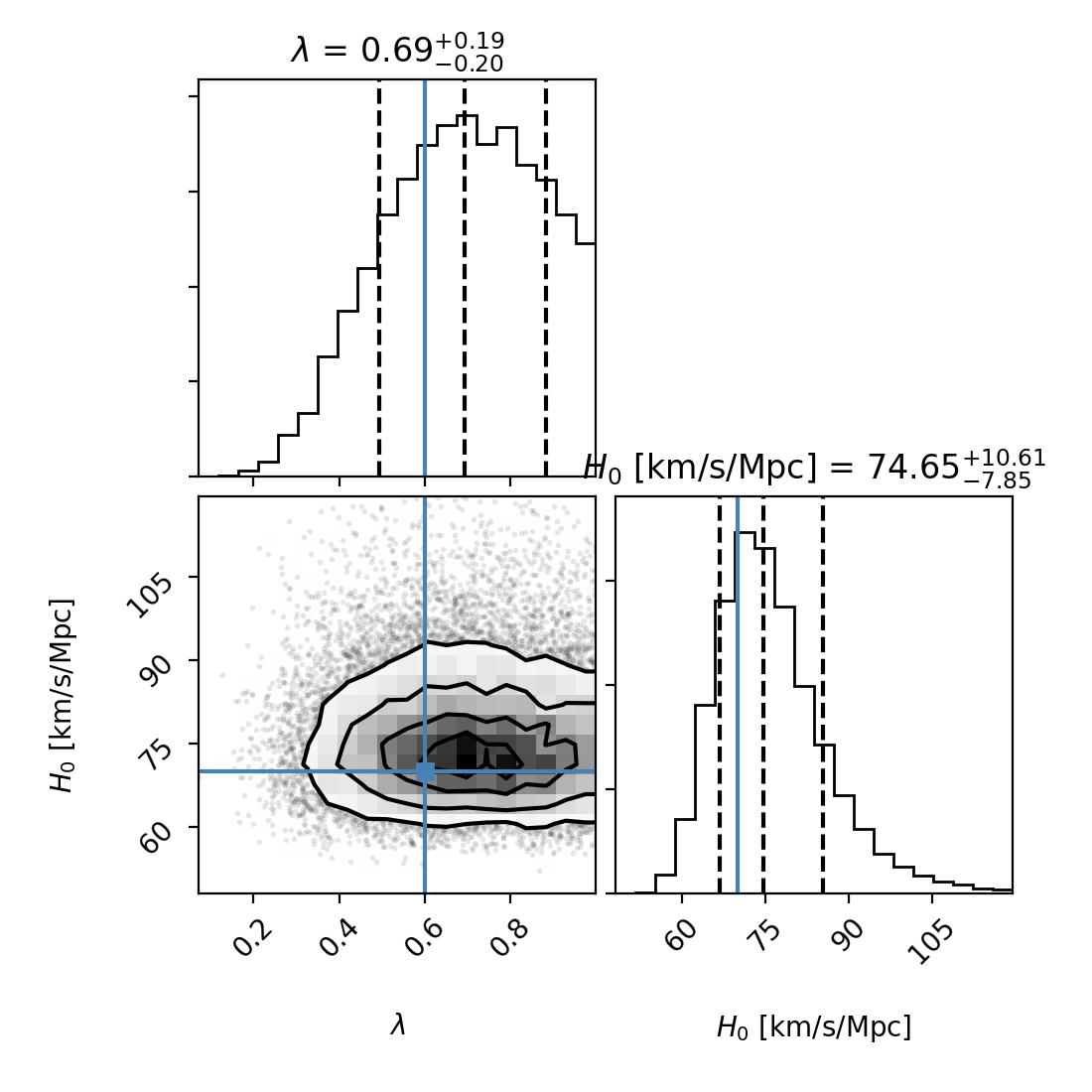}
\caption{Joint posterior of the Hubble constant and the fraction $\lambda$ of BBH giving rise to flares in AGNs from 10 BBH drawn from the top 5\% of the simulated events for LIGO/Virgo/KAGRA O4 with total rest frame mass $>50~M_\odot$. The top events are chosen based on the expected number of background flares, which in turn depends on the localization volume and the flare magnitude (which scales with the total rest-frame BBH mass). The blue square shows the input values of the Hubble constant and of $\lambda$ for the simulation. This figure shows that our statistical framework is able to recover the true values of these parameters, and that the follow-up of 10 well-localized GW BBH can bring a $\sim 12\%$ measurement of $H_0$, provided that $\lambda\sim 0.6$. }\label{fig:lamH0}
\end{figure}

The framework presented in Section \ref{sec:method} is able to provide unbiased constraints on $\lambda$ and cosmological parameters. For this example, we fix the cosmology to a flat $\Lambda$CDM scenario with $\Omega_m=0.3$, and only let $H_0$ vary. We simulate signal and background flares assuming $H_0=70$ km~s$^{-1}$ Mpc$^{-1}$, and we randomly draw 10 events from the top 5\% of the simulated events for LIGO/Virgo/KAGRA O4 with total rest frame mass $>50~M_\odot$, following the simulations described in Section \ref{sec:future}. We assume that 60\% of the events give rise to an AGN flare, which has a mass-dependent magnitude, and hence a mass-dependent rate of background events. Our recovered joint posterior on $\lambda$ and the Hubble constant is shown in Fig.~\ref{fig:lamH0}. This number of well localized and heavy events is expected to be available after $\sim 2$ years of LIGO/Virgo/KAGRA run at the sensitivity expected for O4, and, as already clear from the results of Section \ref{sec:future}, it is expected to place a significant constraint on $\lambda$ as long as $\lambda>0.1$. For a true fraction of BBH in AGNs of $\lambda=0.6$, after marginalizing over the true value of $\lambda$, the expected precision on $H_0$ from 10 events is $\sim 12\%$.

\section{Conclusions}\label{sec:conclusions}

In this paper we show a statistical approach to measure the fraction of GW BBH mergers that induce AGN flares using BBH follow-up observations. First, we show that the AGN flare observed in coincidence with GW190521 is consistent with a background AGN flare, in other words, it is possibly uncorrelated with the GW event as the association cannot be made with confidence. We then show that follow-up campaigns of GW BBH events such as the one performed by ZTF for GW190521 can however effectively constrain the fraction of BBH produced in AGNs, assuming that an electromagnetic counterpart can arise from the BBH, as predicted in \citet{McKernan19}. Assuming that counterpart candidates will be similar to the candidate in \citet{Graham} (or that more generic counterparts occur with a similar frequency in AGNs), and under conservative assumptions about the AGN number density, we find that follow-up campaigns of well-localized BBH mergers will be much more effective at constraining the fraction of BBHs formed in AGNs than methods that do not rely on follow-up observations, and that a confident association could be already possible during the upcoming LIGO/Virgo/KAGRA run, O4. This is particularly important if multiple formation channels contribute to the observed GW BBH mergers \citep{zevin2020channel}, i.e. $\lambda<1$. Even if $\lambda\sim 0.1$, following up $\lesssim 10$ well-localized events will yield an informative measurement, whereas without followup, we would need $10^3- 10^4$ events (which will not be available during O4).

We extend the formalism to measure cosmological parameters in the presence of signal and background flares. We show that this formalism can provide a joint posterior of the Hubble constant and $\lambda$. Assuming a flat $\Lambda$CDM cosmology with $\Omega_m=0.3$, we recover a $\sim 12\%$ precision on the Hubble constant from follow-up observations of 10 well-localized events. Future studies of these sources may also reveal interesting constraints on $\Omega_m$ and the dark energy equation of state, given that the typical distances of the events considered is $\gtrsim 1$ Gpc.

It is worth noting that the SF, which we use here, only quantifies the probability of a quasar luminosity excursion as a Gaussian variance, from which we compute a Gaussian probability. While this is more general than the use of Gaussians in a DRW model, the type of flare expected for this channel may not be well described by a Gaussian process. In the future, it will be interesting to empirically constrain the statistics of the specific flare expected from this BBH merger channel from a large AGN sample as in \citet{graham2017}, and then use that to derive a false-alarm probability and a constraint on $\lambda$. The method presented here is flexible enough so that a change of this kind can be easily incorporated. 

We have applied the method presented to BBH in AGNs for current generation GW detectors, but in the future it could also be interesting to apply this method to other kinds of possible BBH counterparts and to LISA massive black hole binaries (MBH), since similar conditions with several possible variable AGNs in the localization volume may occur.

\acknowledgments
We thank Zoltan Haiman, Imre Bartos, Doga Veske, Paul Martini, Saavik Ford, Robert Morgan, Marica Branchesi, Charlie Kilpartick, and Tamara Davis for very useful discussion on this topic. M.~F. is supported by NASA through NASA Hubble Fellowship grant HST-HF2-51455.001-A awarded by the Space Telescope Science Institute. C.J.B. acknowledges support from the Illinois Graduate Survey Science Fellowship.

Work supported by the Fermi National Accelerator Laboratory, managed and operated by Fermi Research Alliance, LLC under Contract No. DE-AC02-07CH11359 with the U.S. Department of Energy. The U.S. Government retains and the publisher, by accepting the article for publication, acknowledges that the U.S. Government retains a non-exclusive, paid-up, irrevocable, world-wide license to publish or reproduce the published form of this manuscript, or allow others to do so, for U.S. Government purposes.

\bibliographystyle{yahapj}
\bibliography{references}

\end{document}